\documentclass[a4paper,11pt]{article}
\usepackage{pos}

\title{\texttt{ggxy}: NLO QCD corrections to loop induced $gg$ initiated processes}

\author*[a]{Daniel Stremmer}

\affiliation[a]{ Institut f{\"u}r Theoretische Teilchenphysik, Karlsruhe Institute of Technology (KIT), \\Wolfgang-Gaede Stra\ss{}e 1, 76131 Karlsruhe, Germany}

\emailAdd{daniel.stremmer@kit.edu}

\abstract{
We present the program package \texttt{ggxy}, which in its first version can be used to calculate partonic and hadronic cross sections to Higgs boson pair production at NLO QCD and it offers a direct interface to \texttt{POWHEG-BOX}. The 2-loop virtual amplitudes are implemented using analytical approximations in different kinematic regions, while all other parts of the calculation are exact. This implementation allows to freely modify all input parameter, such as the three Higgs coupling, the masses of the top quark and the Higgs boson, and the top-quark mass renormalization scheme.

\bigskip

P3H-25-097, \, TTP-25-050
}

\FullConference{The European Physical Society Conference on High Energy Physics (EPS-HEP2025)\\
7-11 July 2025\\
Marseille, France\\}

\begin{document}
\maketitle

\section{Introduction}
One of the goals of the High-Luminosity LHC (HL-LHC) is the observation of Higgs pair production which can be directly related into a measurement of the three Higgs coupling ($\lambda_3$) and provides an important consistency check of the Standard Model (SM). Recent projections \cite{CMS:2025hfp} show that an observation of this process is indeed expected from the combined ATLAS+CMS measurement at $3ab^{-1}$ with a significance of $7\sigma$. This translates to a measurement of $\lambda_3$ with a precision of at least $30\%$ assuming no BSM effects.

Higgs pair production in hadron collisions, such as in the LHC, is dominated by the loop-induced gluon fusion process for which the LO results are well known for many years \cite{Glover:1987nx,Plehn:1996wb}. The calculation of NLO QCD corrections with exact top-quark mass dependence has been obtained for the first time in Refs. ~\cite{Borowka:2016ehy,Borowka:2016ypz}, where the two-loop integrals have been calculated numerically using (Py-)SecDec \cite{Carter:2010hi,Borowka:2017idc}. However, this approach suffered from a long runtime since the numerical integration of the two-loop integrals has to be redone for every phase-space point. In order to avoid this problem, an interpolation grid ({\tt hhgrid}) \footnote{\url{https://github.com/mppmu/hhgrid}} for the virtual correction has been generated. While such an grid provides a very fast evaluation time, there are still drawbacks since the numerical values of $m_H$ and $m_t$ has been fixed and thus cannot be varied. The latter also prohibits the use of this grid for calculations in the $\overline{\rm MS}$ top-quark mass scheme. However, such a study on the top-quark mass scheme dependence has been performed in Ref. \cite{Baglio:2018lrj}, where again the two-loop integrals have been calculated using numerical approaches, where no (exact) timings have been presented.

One way to avoid the problems of a long runtime and the limited flexibility is to use analytical approximations in specific kinematical regions for the computation of the virtual corrections. In particular, it was shown that the combination of the high-energy expansion \cite{Davies:2018ood,Davies:2018qvx} and the expansion around small transverse Higgs boson momenta \cite{Bellafronte:2022jmo,Davies:2023vmj} can be combined in order to cover the full phase space.

In this proceeding, we present the flexible \texttt{C++} library \texttt{ggxy} \cite{Davies:2025qjr}, which provides in its first version all necessary functions for the computation of partonic form factors and hadronic cross sections for $gg\to HH$ at NLO QCD while keeping the full Higgs and top-quark mass dependence. The two-loop amplitudes are implemented using analytical approximations \cite{Davies:2018ood,Davies:2018qvx,Davies:2023vmj}, while all other parts are kept exact. In addition, the library provides a good computational performance, with typical runtimes of $30$ mins for the computation of total hadronic cross sections with a statistical precision of less than $0.2\%$. Finally, an interface to \texttt{POWHEG-BOX}~\cite{Alioli:2010xd} is available which can be used for the matching to Parton showers.

\section{\texttt{ggxy} library}
The main advantage of \texttt{ggxy} with respect to purely numerical approaches for the computation of the virtual corrections is the use of analytical approximation for them. This allows a fast runtime and enables a high flexibility in the variation of input parameters. In particular, we have implemented in \texttt{ggxy} the analytical expressions for the form factors $F_1$ and $F_2$ for $gg\to HH$ based on the high-energy \cite{Davies:2018ood,Davies:2018qvx} and the forward expansions \cite{Davies:2023vmj}. These approximations have been obtained by considering different kinematic hierarchies of the scales involved in the process. In particular, the expressions of the forward expansion have been obtained by considering $m_H^2,|t|\ll m_t^2,s$, while the high-energy expansion is calculated by employing $m_H^2\ll m_t^2\ll s,|t|$. In addition, the region of convergence of the high-energy expansion is further increased by the construction of a Pad\'e approximant. In practise we use the expressions from the forward expansion for $p_T<200$~GeV and from the high-energy expansion for $p_T>220$~GeV. The results in the intermediate region are obtained from an interpolation of both approximations. More details can be found in Ref. \cite{Davies:2025qjr}.

We use {\tt Recola}~\cite{Actis:2016mpe} and {\tt Collier}~\cite{Denner:2016kdg} for the computation of the $2\to 3$ matrix elements required for the real corrections, with {\tt CutTools}~\cite{Ossola:2007ax} and {\tt OneLOop}~\cite{vanHameren:2010cp} as a fallback option for the reduction to and evaluation of scalar integrals as an additional cross-check for exceptional phase-space points. IR singularities are subtracted with the Catani-Seymour dipole subtraction scheme \cite{Catani:1996vz}. The phase-space integration is handled with the help of {\tt avhlib}~\cite{vanHameren:2007pt,vanHameren:2010gg} and we use {\tt CRunDec}~\cite{Herren:2017osy} for the running and decoupling of $\alpha_s$ entering the running of the top-quark mass in $\overline{\rm MS}$ scheme, which is also done with {\tt CRunDec}. In addition, we use the code of Ref. \cite{Frellesvig:2016ske} for the evaluation of polylogarithmic functions.

\texttt{ggxy} can be obtained from the repository
\begin{equation*}
\textrm{\url{https://gitlab.com/ggxy/ggxy-release}}
\end{equation*}
and is directly shipped together with the tools mentioned above. The installation is handled with \verb|CMake| and the following external dependencies has to be installed first: {\tt boost}\footnote{\url{https://www.boost.org}}, {\tt eigen}\footnote{\url{https://libeigen.gitlab.io/docs/}} and {\tt LHAPDF}~\cite{Buckley:2014ana}.

\section{Example results}

\begin{table}[t]
    \centering
    \renewcommand{\arraystretch}{1.2}
    \begin{tabular}{cc@{\hskip 10mm}l@{\hskip 10mm}l@{\hskip 10mm}}
        \hline
        $\sqrt{s}$&
        &{ \tt ggxy}&Ref.~\cite{Borowka:2016ypz}  \\
        \hline
        14~TeV & $\sigma^{\rm LO}$ [fb]& $19.848(4)^{+27.6\%}_{-20.5\%}$ &    $19.85^{+27.6\%}_{-20.5\%}$\\
        & $\sigma^{\rm NLO}$ [fb]& $32.92(2)^{+13.6\%}_{-12.6\%}$ & $32.91^{+13.6\%}_{-12.6\%}$\\
        \noalign{\smallskip}\hline\noalign{\smallskip}
        100~TeV & $\sigma^{\rm LO}$ [fb]& $731.2(2)^{+20.9\%}_{-15.9\%}$ &    $731.3^{+20.9\%}_{-15.9\%}$\\
        & $\sigma^{\rm NLO}$ [fb]& $1150(1)^{+10.8\%}_{-10.0\%}$ & $1149^{+10.8\%}_{-10.0\%}$\\        
        \noalign{\smallskip}\hline\noalign{\smallskip}
    \end{tabular}
    \caption{\label{tab::sig}Comparison with results of Ref.~\cite{Borowka:2016ypz}
    for $\sqrt{s}=14$ and $\sqrt{s}=100$~TeV. Table taken from Ref. \cite{Davies:2025qjr}.}
\end{table}

Examples for the usage of \texttt{ggxy} can be found in the subdirectories \verb|examples/gghh-FF/| and \verb|examples/gghh-nlo/| for the computation of partonic form factors/matrix elements and hadronic cross sections, respectively. In particular, we provide in the latter subdirectory the example program \verb|nlo-gghh.cpp|, which can be used to compute integrated and differential cross section at LO and NLO QCD. The runtime of the NLO QCD computation takes about $30$ mins on a single core which leads to statistical uncertainties of less than $0.2\%$ for the total hadronic cross section. The default parameters are adapted from Ref.~\cite{Borowka:2016ypz}, so that \texttt{ggxy} directly reproduces the LO and NLO cross sections at $14$ and $100$ TeV as shown in Table \ref{tab::sig}.

\begin{figure}[t]
  \begin{center}
  \begin{tabular}{cc}
     \includegraphics[width=0.45\textwidth]{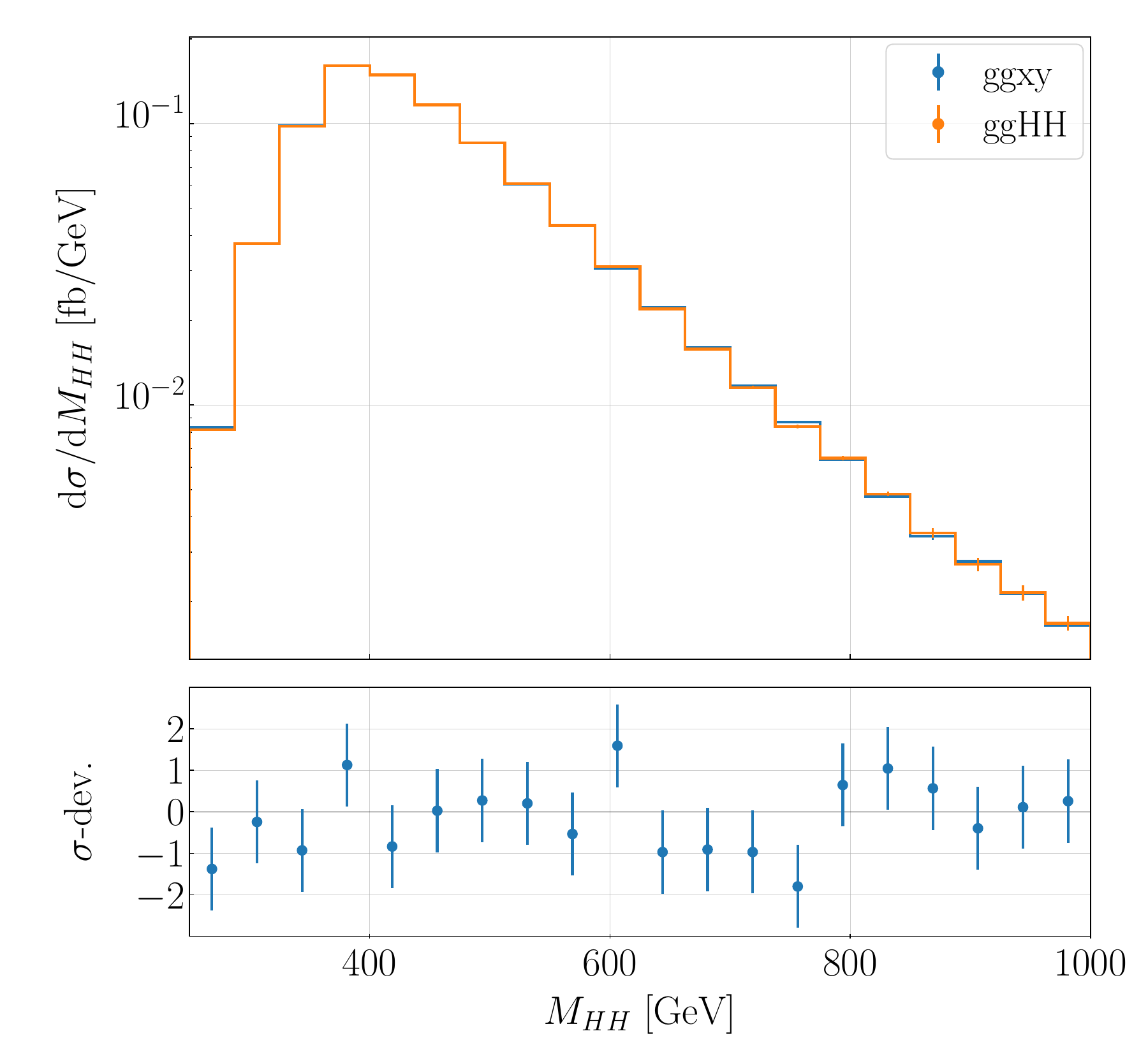}
     \includegraphics[width=0.45\textwidth]{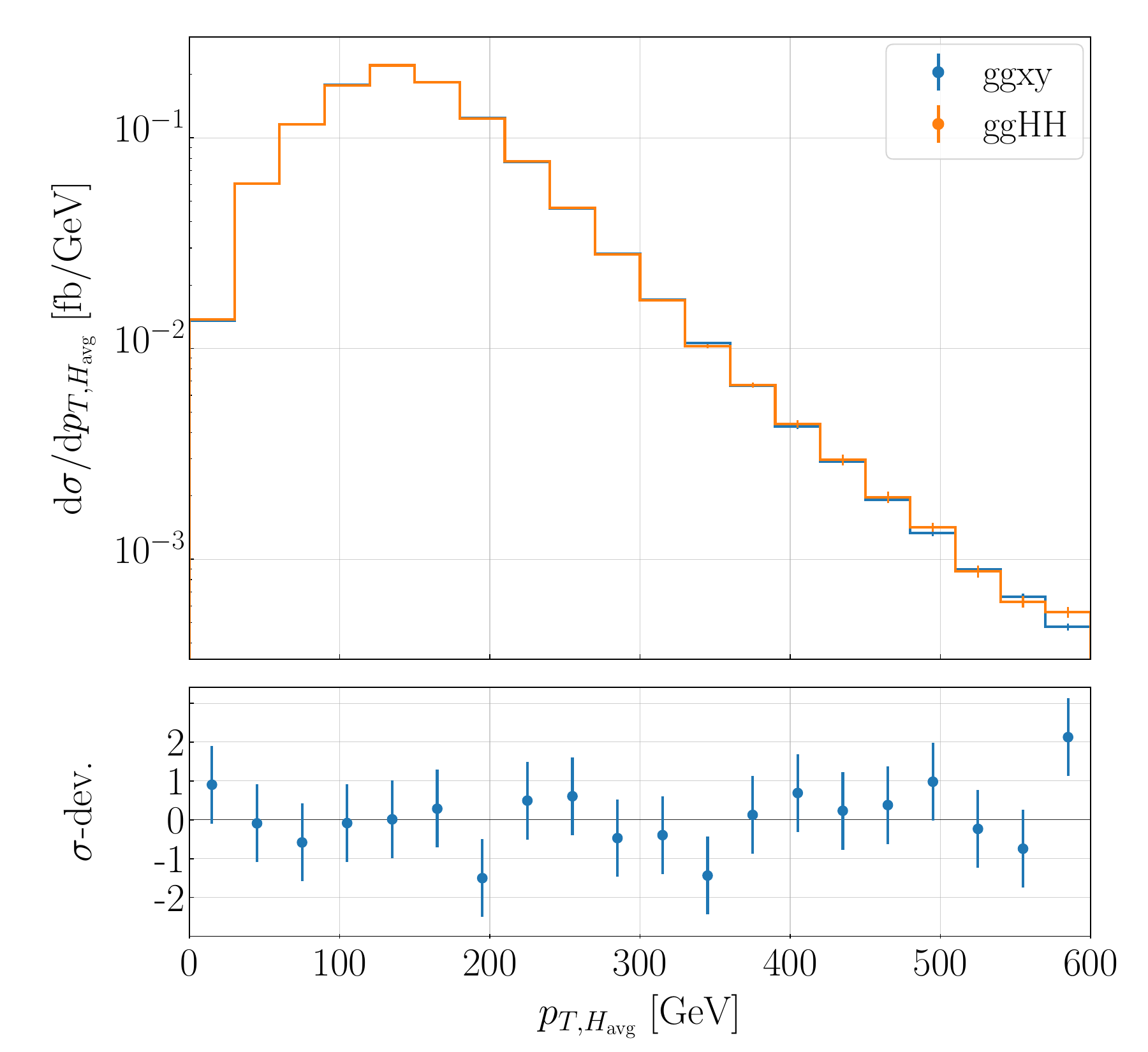}
  \end{tabular}
  \end{center}
  \caption{\label{fig::nlo} Differential cross-section distributions for the observables $m_{HH}$ and $p_{T,H_{\rm avg}}$ at NLO QCD and $\sqrt{s}=14$~TeV computed with the POWHEG implementations \texttt{ggxy\_ggHH} and \texttt{ggHH} \cite{Heinrich:2017kxx}. Lower panels display $\sigma$ deviations with respect to the statistical uncertainties.}
\end{figure}

In addition, a user process for $gg\to HH$ based on \texttt{ggxy} and \texttt{Recola} has been implemented in \texttt{Powheg}~\cite{Alioli:2010xd}, which can be obtained from
\begin{equation*}
\textrm{\url{https://gitlab.com/POWHEG-BOX/V2/User-Processes/ggxy_ggHH}}.
\end{equation*}
The runtime of this implementation is faster by a factor of $4-5$ compared to the previous implementation of $gg\to HH$ in \texttt{Powheg} (\texttt{ggHH}) \cite{Heinrich:2017kxx} based on {\tt hhgrid}, which is mainly caused by the fast numerical evaluation of the one-loop $2\to 3$ matrix elements for the real corrections with \texttt{Recola}. A comparison of both implementations at the differential level for the observables $m_{HH}$ and $p_{T,H_{\rm avg}}$ at NLO QCD is shown in Figure \ref{fig::nlo}. The lower panels display the $\sigma$ differences between both calculations in standard deviations (denoted as $\sigma$-dev.), which indicate excellent agreement for the fixed values of $m_t$ and $m_H$ in {\tt ggHH}.

\section{Conclusions and outlook}
In this proceeding we have presented the new library \texttt{ggxy} which can be used for the calculation of NLO QCD correction to Higgs pair production in gluon fusion with full top-quark mass dependence. The virtual corrections have been implemented using analytical approximations which makes the evaluation fast and flexible, so that all input parameters, such as $m_H$, $m_t$, $\kappa_\lambda=\lambda_3/\lambda_3^{\rm SM}$ and the top-quark mass scheme, can be freely chosen. In addition, \texttt{ggxy} has been interfaced to \texttt{POWHEG-BOX}, which allows the matching to parton showers while keeping the flexibility to vary all input parameters.

The methods used for $gg\to HH$ can also directly be applied on other processes such as $gg\to ZH$ or top-quark mediated off-shell $gg\to Z^\star Z^\star$ for which the analytical approximations of the two-loop amplitudes have been computed in Ref. \cite{Davies:2025out}. The expressions of $gg\to Z^\star Z^\star$, can also be used to easily the extract the amplitudes of top-quark mediated $gg\to \gamma^\star\gamma^\star$ and $gg\to Z^\star\gamma$. These processes will become available in a future version of \texttt{ggxy}.

\section*{Acknowledgments}
This research was supported by the Deutsche Forschungsgemeinschaft (DFG, German
Research Foundation) under grant 396021762 — TRR 257: {\it P3H - Particle Physics Phenomenology after the Higgs Discovery}.


\bibliographystyle{JHEP}
\bibliography{References.bib,extra.bib}

\end{document}